\documentclass[twocolumn,prb,amsmath,amssymb,floatfix,superscriptaddress,showpacs,longbibliography]{revtex4-1}
\usepackage{graphicx}
\usepackage{dcolumn}
\usepackage{bm}
\usepackage{xspace}
\usepackage{hyperref}
\usepackage{color}
\def\new{\color{black}}
\def\cto{Cu$_{3}$TeO$_{6}$\xspace}

\begin{document}

\title{Evidence for magnon-phonon coupling in the topological magnet \cto}
\author{Song~Bao}
\altaffiliation{These authors contributed equally to the work.}
\author{Zhengwei~Cai}
\altaffiliation{These authors contributed equally to the work.}
\author{Wenda~Si}
\author{Wei~Wang}
\author{Xiaomeng~Wang}
\author{Yanyan~Shangguan}
\author{Zhen~Ma}
\affiliation{National Laboratory of Solid State Microstructures and Department of Physics, Nanjing University, Nanjing 210093, China}
\author{Zhao-Yang~Dong}
\affiliation{Department of Applied Physics, Nanjing University of Science and Technology, Nanjing 210094, China}
\author{Ryoichi~Kajimoto}
\affiliation{J-PARC Center, Japan Atomic Energy Agency (JAEA), Tokai, Ibaraki 319-1195, Japan}
\author{Kazuhiko~Ikeuchi}
\affiliation{Neutron Science and Technology Center, Comprehensive Research Organization for Science and Society (CROSS), Tokai, Ibaraki 319-1106, Japan}
\author{Shun-Li~Yu}
\author{Jian~Sun}
\author{Jian-Xin~Li}
\author{Jinsheng~Wen}
\email{jwen@nju.edu.cn}
\affiliation{National Laboratory of Solid State Microstructures and Department of Physics, Nanjing University, Nanjing 210093, China}
\affiliation{Collaborative Innovation Center of Advanced Microstructures, Nanjing University, Nanjing 210093, China}

\begin{abstract}

We perform thermodynamic and inelastic neutron scattering (INS) measurements to study the lattice dynamics (phonons) of a cubic collinear antiferromagnet \cto which hosts topological spin excitations (magnons). While the specific heat and thermal conductivity results show that the thermal transport is dominated by phonons, the deviation of the thermal conductivity from a pure phononic model indicates that there is a strong coupling between magnons and phonons. In the INS measurements, we find a mode in the excitation spectra at 4.5~K, which exhibits a slight downward dispersion around the Brillouin zone center.  This mode disappears above the N\'{e}el temperature, and thus cannot be a phonon. Furthermore, the dispersion is distinct from that of a magnon. Instead, it can be explained by the magnon-polaron mode, which is new collective excitations resulting from the hybridization between magnons and phonons. We consider the suppression of the thermal conductivity and emergence of the magnon-polaron mode to be evidence for magnon-phonon coupling in \cto.
 
\end{abstract}

\maketitle
\section{Introduction}

Magnons and phonons, which are quanta of spin waves and lattice vibrations, respectively, constitute two fundamental collective excitations in condensed matter physics. They coexist and propagate in ordered magnets, while in the case where magnetoelastic effect is strong, they are coupled and exhibit rich physics. For example, previous studies identified several spectroscopic signatures that could be attributed to magnon-phonon coupling, such as renormalization of spin-wave excitations\cite{PhysRevB.97.201113,PhysRevLett.111.257202,PhysRevB.76.054431}, enhancement of spontaneous magnon decay\cite{RevModPhys.85.219,oh2016spontaneous}, and gap opening at the intersection of magnon and phonon dispersions\cite{PhysRevLett.99.266604,PhysRevB.97.134304}. Besides the direct impacts on excitation spectra, magnon-phonon coupling can also be reflected in some thermal transport measurements, as it can give rise to the suppression of thermal conductivity\cite{PhysRevLett.93.177202,PhysRevB.8.2130} and anomalies in the magnetic-field-dependent spin Seebeck effect\cite{PhysRevLett.117.207203,PhysRevB.95.144420}. More intriguingly, the strong magnetoelastic coupling between magnons and phonons is believed to result in new hybrid quasiparticle excitations, such as electromagnons which are electro-active magnons in multiferroic materials\cite{pimenov2006,PhysRevLett.98.027202,PhysRevB.77.014438,Pimenov_2008,nm9_975,toth2016electromagnon,PhysRevB.96.041117}, and magnon polarons which are hybridized magnon and phonon modes in proximity of the intersection of the uncoupled magnon and phonon dispersions\cite{RevModPhys.21.541,PhysRevB.91.104409,PhysRevLett.115.197201,PhysRevLett.117.207203,PhysRevB.95.144420,PhysRevB.96.104441,PhysRevLett.121.237202,PhysRevB.99.064421,PhysRevB.99.214445}. More recently, magnon polarons have been discussed in the context of topology\cite{PhysRevLett.117.217205,PhysRevLett.122.107201,PhysRevB.99.174435,PhysRevLett.123.237207}. In particular, the magnon-polaron bands were proposed to carry Chern numbers and exhibit large Berry curvature\cite{PhysRevLett.117.217205,PhysRevLett.122.107201,PhysRevB.99.174435,PhysRevLett.123.237207}. Such proposals provide a new platform to study topological bosonic excitations\cite{PhysRevLett.115.147201,McClarty2017,PhysRevX.8.041028,Nawa2019} and the associated thermal Hall effect\cite{Ideue2017,PhysRevLett.123.167202,park2019thermal} in real magnets. Besides the fundamental importance of the magnon-phonon coupling, it also holds promising application potentials such as in developing multifunctional devices and low-cost dispationless spintronics\cite{nature464_262,nm9_975,doi:10.1063/1.3624845,Tokura_2014,uchida2011,PhysRevLett.108.176601,PhysRevLett.121.237202,Ogawa8977}. As such, observing the magnon-phonon coupling experimentally and understanding the consequent phenomena is one of the central topics in condensed matter physics. 

In ferrimagnet yttrium iron garnet\cite{holanda2018detecting,PhysRevLett.117.207203,PhysRevB.96.100406} and non-collinear antiferromagnets including hexagonal rare-earth manganites $R$MnO$_{3}$ family ($R$=Y, Lu, Ho, Yb, Sc, Tm, Er) (Refs~\onlinecite{Sim:bm5079,PhysRevLett.111.257202,PhysRevLett.99.266604,oh2016spontaneous,PhysRevB.97.201113,PhysRevB.97.134304,lee2008giant,PhysRevLett.93.177202,C5TC02096D}), BiFeO$_{3}$ (Ref.~\onlinecite{PhysRevB.91.064301}), CuCrO$_{3}$ (Ref.~\onlinecite{PhysRevB.94.104421}), LiCrO$_{3}$ (Ref.~\onlinecite{toth2016electromagnon}), CuBr$_{2}$ (Ref.~\onlinecite{PhysRevB.96.085111}) and Mn$_{3}$Ge (Ref.~\onlinecite{PhysRevB.99.214445}), magnon-phonon coupling has been heavily explored. On the other hand, 
the coupling mechanism in collinear antiferromagnet remains elusive and calls for experimental investigations\cite{PhysRevB.99.064421}. In this paper, we study the coupling in \cto. It develops a long-range collinear antiferromagnetic order below the N\'{e}el temperature $T_{\rm N}$ of 61~K, with spins aligned along the [111] direction of the cubic unit cell\cite{herak2005novel}. It was reported that there were phonon anomalies in \cto, evidenced by the emergence of new modes below 50~K observed by Raman scattering\cite{Choi2008} and optical reflectivity measurements\cite{Caimi_2006}. It was proposed that the phonon anomalies could be explained by a magnetoelastic strain induced by the collinear antiferromagnetic order\cite{Choi2008,Caimi_2006}. Furthermore, such a magnetic structure also preserves the $PT$ symmetry, where $P$ and $T$ are space-inversion and time-reversal operations, respectively\cite{PhysRevLett.119.247202}. Previous studies\cite{PhysRevLett.119.247202,PhysRevB.99.035160,Bao2018,Yao2018} demonstrated that under the protection of the $PT$ symmetry, \cto could host topological magnons, as observed directly in the magnetic excitation spectra using INS\cite{Bao2018,Yao2018}. Because of the $PT$ symmetry, both the electronic and phononic bands can host topological band structures\cite{Bradlynaaf5037,nature547_298}. Considering these, \cto provides an ideal platform to investigate the interplay between topological magnons and phonons in a three-dimensional collinear antiferromagnetic case.

Here, we report comprehensive results of specific heat, thermal conductivity and INS measurements on \cto. We find that the specific heat and the thermal conductivity are mostly contributed by phonons and the magnetic contributions are negligible. However, our measured thermal conductivity shows an obvious suppression due to the magnon-phonon coupling. In our INS measurements, we observe an anomalous mode located around 16.8~meV at 4.5~K, which disappears above $T_{\rm N}$ and is believed to be a magnon-polaron mode due to the hybridization between magnons and phonons. Our results on the suppression of the thermal conductivity and emergence of the magnon polarons are compelling evidence for the magnon-phonon coupling in \cto.  

\section{Experimental Details}

Single crystals of Cu$_{3}$TeO$_{6}$ were grown by the flux method as described in Ref.~\onlinecite{HE2014146}. Specific heat and thermal conductivity were measured using the Heat Capacity Option and the Thermal Transport Option, respectively, integrated in a Physical Property Measurement System (PPMS) from Quantum Design. Neutron scattering measurements were performed on 4D-Space Access Neutron Spectrometer (4SEASONS) at Materials and Life Science Experimental Facility (MLF) of Japan Proton Accelerator Research Complex (J-PARC)\cite{doi:10.1143/JPSJS.80SB.SB025}. For the experiment on 4SEASONS, 50 pieces of single crystals weighing about 6.3~g in total were coaligned and glued on two aluminum plates, using a backscattering Laue X-ray diffractometer. The crystals were well coaligned with an overall sample mosaic of $3^{\circ}$. The assembly was mounted in a closed-cycle refrigerator with the [010] direction along the vertical direction and [100] along the incident neutron beam direction. We used a primary incident energy $E_{i}=40$~meV and a chopper frequency of 250~Hz, resulting in an energy resolution of 2.4~meV at the elastic line. Measurements were done at 4.5 and 100 K. We set the angle where the [100] direction is parallel to the incident beam direction to be 0$^{\circ}$. Data were collected by rotating the sample about the [010] axis from $30^{\circ}$ to $90^{\circ}$ in a $0.25^{\circ}$ step. We counted 18 minutes for each step. Raw data were reduced and combined together into four-dimensional matrices, and analyzed using the software suite Utsusemi\cite{inamura2013development}. We used the corrected wave vector (1,\,-0.0225,\,0) instead of (1,\,0,\,0) to represent the direction parallel to neutron beam to account for the small sample tilt. After the correction, all Bragg peaks were located well at the reciprocal lattice points. The neutron scattering data were described in reciprocal lattice unit (rlu) of $(H,\,K,\,L)$=$(2\pi/a,\,2\pi/b,\,2\pi/c)$ with $a=b=c=9.537(3)$~\AA.

\begin{figure}[htb]
\centering
\includegraphics[width=0.9\linewidth]{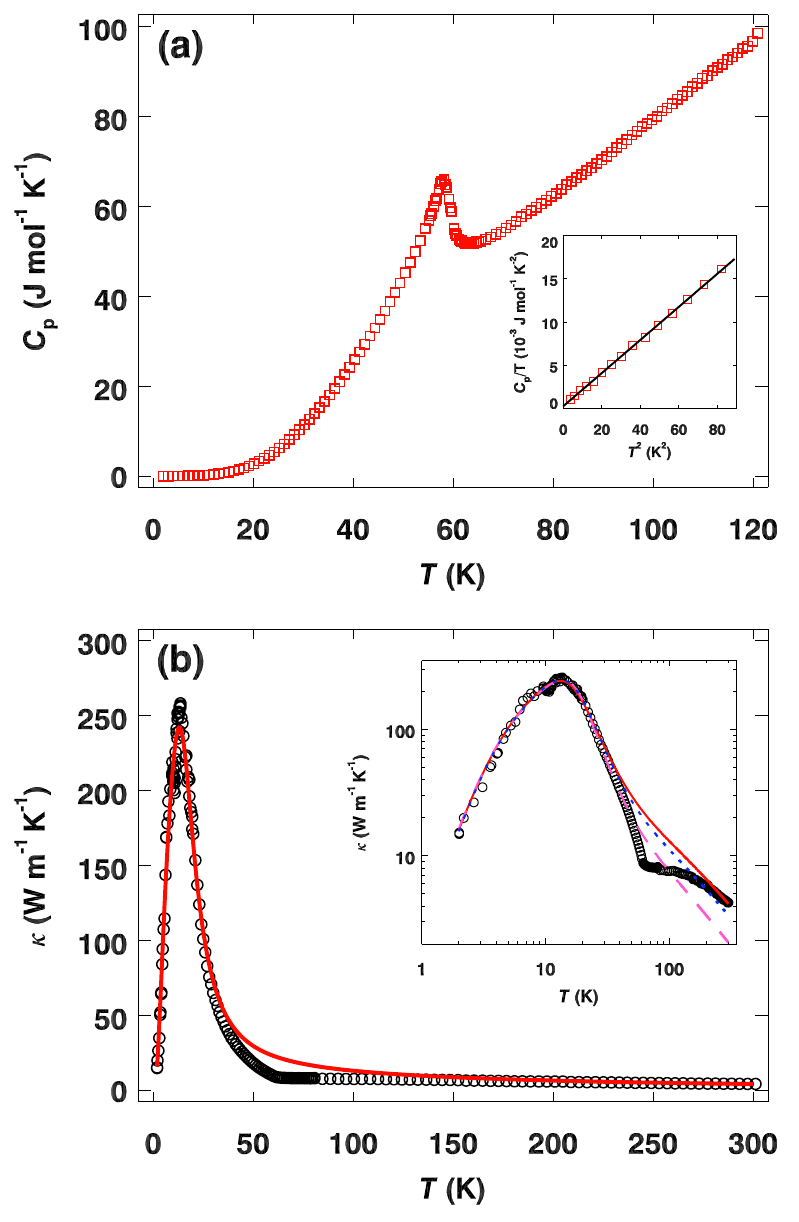}
\caption{\label{fig:thermal}{(a) Temperature dependence of the specific heat for Cu$_{3}$TeO$_{6}$. The inset is the low-temperature specific heat plotted as $C_{\rm {p}}/T$ vs $T^{2}$. The solid line is a linear fit. (b) Thermal conductivity measured along one of the three principal crystallographic axes. The inset is the measured data plotted in logarithmic scale. {\new Curves represent different fits as described in the main text.}}}
\end{figure}

\section{Results}

\subsection{Thermodynamic Measurements}

Results of the thermodynamic measurements including specific heat and thermal conductivity on Cu$_{3}$TeO$_{6}$ are shown in Fig.~\ref{fig:thermal}. The antiferromagnetic phase transition temperature $T_{\rm N}$ of 61~K is indicated by both the $\lambda$-type peak in the specific heat [Fig.~\ref{fig:thermal}(a)] and the kink in the thermal conductivity [Fig.~\ref{fig:thermal}(b)]. Since Cu$_{3}$TeO$_{6}$ is a magnetic insulator, only lattice vibrations and magnetic excitations can contribute the low-temperature specific heat, which can be written as\cite{Li2002} 
\begin{equation}\label{HC}
  C_{\rm {p}}(T)=\frac{12}{5}\pi^{4}xN_{\rm A}k_{\rm B}(\frac{T}{\Theta_{\rm D}})^3+6.8yN_{\rm A}k_{\rm B}(\frac{k_{\rm B}T}{12|J|S})^3,
\end{equation}
where $N_{\rm A}$, $k_{\rm B}$, $T$, $\Theta_{\rm D}$, $J$ and $S$ are Avogadro constant, Boltzmann constant, temperature, Debye temperature, antiferromagneitc exchange interaction and spin, respectively; $x$ and $y$ represent the number of total atoms and magnetic ions per formula unit, respectively. The first and second term in Eq.~\ref{HC} correspond to the specific heat contributed by phonons ($C_{\rm ph}$) and magnons ($C_{\rm m}$), respectively .

From Eq.~\ref{HC}, we find it difficult to experimentally distinguish the lattice and the magnetic contributions from the total specific heat, because they both follow the $T^{3}$ power-law behaviors at low temperatures. However, if we take $J$ of~$\sim11$~meV and $S$ of 1/2 for \cto
(Ref.~\onlinecite{Bao2018}) in Eq.~\ref{HC}, the magnetic specific heat $C_{\rm m}$ is at least three orders of magnitude smaller than $C_{\rm ph}$. Therefore, we simplify the low-temperature specific heat as $C_{\rm ph}$. In the inset of Fig.~\ref{fig:thermal}(a), we plot  $C_{\rm p}/T$ as a function of $T^2$. It can be nicely fitted with a straight line having a slope of $1.91\times10^{-4}~{\rm J}~{\rm mol}^{-1}~{\rm K}^{-4}$, from which we deduce the Debye temperature $\Theta_{\rm D}~\sim466$~K. 

The temperature evolution of the thermal conductivity $\kappa$ for Cu$_{3}$TeO$_{6}$ is shown in Fig.~\ref{fig:thermal}(b). As cooling from room temperature, $\kappa$ increases slowly in the beginning until reaching the $T_{\rm N}$. Below this point, it increases dramatically and reaches its maximum value of about $250~{\rm W}~{\rm m}^{-1}~{\rm K}^{-1}$  and then drops rapidly, forming a peak at about 14~K. Since $\kappa$ is proportional to the specific heat, it should be dominated by phonons as $C_{\rm {ph}}$ is much larger than $C_{\rm m}$. Indeed, the observed low-temperature peak is one of the typical characteristics for phonon thermal conductivity ($\kappa_{\rm ph}$) (Refs~\onlinecite{PhysRev.126.427,PhysRevB.4.592,Hess2007}). The behavior of $\kappa_{\rm ph}$ is determined by various scattering mechanisms. Generally speaking, in an ideal phononic crystal without any defects, in the high-temperature limit, the thermal conductivity usually exhibits $T^{-1}$ behavior since the phonon-phonon umklapp process dominates and phonon mean-free path increases as temperature decreases; while in the low-temperature limit, the dominating scattering process is only related to the sample size, causing a temperature-independent phonon mean-free path, so the thermal conductivity exhibits $T^{3}$ behavior as the specific heat does\cite{PhysRev.126.427,PhysRevB.4.592,Hess2007}. The different temperature dependences resulting from these two scattering mechanisms are expected to lead to a typical phonon-peak at low temperatures\cite{PhysRev.126.427,PhysRevB.4.592,Hess2007}. 

However, in a real material, defects and impurities are inevitable and can also scatter phonons. Moreover, magnons start to establish below $T_{\rm N}$ in \cto that can affect the thermal conductivity by acting either as the carriers of heat current or scatterers of phonons\cite{PhysRevLett.95.156603}. In order to understand the thermal transport processes and figure out the scattering mechanisms in Cu$_{3}$TeO$_{6}$, the Debye-Callaway model\cite{Berman1976,PhysRev.113.1046}, which is applicable to $\kappa_{\rm ph}$ calculations, is used to fit our data. In this model, different scattering mechanisms can be reflected by the different terms in the relaxation rates. With this model, $\kappa_{\rm ph}$ can be written as\cite{Berman1976,PhysRev.113.1046} 
\begin{equation}\label{kappa}
  \kappa_{\rm ph}=\frac{k_{\rm B}}{2\pi^{2}v}(\frac{k_{\rm B}T}{\hbar})^{3}\int_{0}^{\frac{\Theta_{\rm D}}{T}}\frac{z^{4}{\rm e}^{z}}{({\rm e}^{z}-1)^2}\tau(\omega,T)dz,
\end{equation}
where $\hbar$, $\omega$, and $v$ are reduced Planck constant, phonon frequency, and phonon mean velocity, respectively, and $z$ is defined as $\hbar\omega/k_{\rm B}T$. Here, with the Debye temperature $\Theta_{\rm D}$ of 466~K, $v$ is calculated to be about 3900~m/s, using the equation $v=\Theta_{\rm D}(k_{\rm B}/\hbar)(6\pi^2n)^{-1/3}$, where $n$ is the number of atoms per unit volume. The relaxation time $\tau(\omega,T)$ in Eq.~\ref{kappa} can be approximated by
\begin{equation}\label{tau} 
  \tau^{-1}(\omega,T)=\frac{v}{L}+A\omega^{4}+BT\omega^{2}{\rm exp}(-\frac{\Theta_{\rm D}}{bT}).
\end{equation}
Here, the three terms correspond to phonon scattering by sample boundaries, point defects and phonon-phonon U processes, respectively\cite{PhysRev.133.A253,PhysRevB.64.054412}. $L$, $A$, $B$, and $b$ are free parameters.

To fit our data in Fig.~\ref{fig:thermal}(b), the Levenberg-Marquardt algorithm with the least-square criterion\cite{Marquardt1963,10.1007/BFb0067700} was used. {\new From the fitting, shown as the dashed curve in the inset of Fig.~\ref{fig:thermal}(b), we obtain the parameters $L=0.93$~mm, $A=2.04\times10^{-43}$~s$^{3}$, $B=2.03\times10^{-17}$~s~K$^{-1}$ and $b=4.68$. This fitting agrees with the low-temperature data well but does not capture the magnetic transition at $T_{\rm N}$, and more worse, deviates from the data significantly above 100~K. Since phonon-phonon umklapp processes are dominant for temperatures far above the magnetic transition temperature, such a deviation at high temperatures is unexpected. Since around $T_{\rm N}$ there may be some other scattering mechanisms, such as critical scattering or magnon-phonon coupling that may affect the fitting, we fit the data in the range away from $T_{\rm N}$, {\it i.e.}, at low (2-25~K) and high temperatures (150-300~K), and obtain the parameters as $L=0.86$~mm, $A=1.88 \times 10^{-43}$~s$^3$, $B=8.93 \times 10^{-18}$~s~K$^{-1}$ and $b=5.64$. The parameter $L$ is comparable with the expected value of $2\sqrt{S_{\rm CS}/\pi}=2.0$~mm, where $S_{\rm CS}$ represents the rectangular cross section of the sample. The parameter $b$ is close to the expected value of 2$\nu^{1/3} = 6.84$, where $\nu$ is the number of atoms in the primitive unit cell\cite{PhysRevB.64.054412}. The fit shown as the solid curve in Fig.~\ref{fig:thermal}(b) matches the experimental data well except for the temperatures ranging from 40 to 200~K. As an attempt to match the data around $T_{\rm N}$ better, we fix the values of $L=0.86$~mm and $A=1.88 \times 10^{-43}$~s$^3$, and refit the data in the whole temperature range with only $B$ and $b$ as tuning parameters. It returns the values of $B$ and $b$ as $1.15 \times 10^{-17}$~s~K$^{-1}$ and 5.33, respectively. However, this new fitting, shown as the dotted line in the inset of Fig.~\ref{fig:thermal}(b), does not improve the fitting around $T_{\rm N}$ significantly, but causes an unexpected deviation at high temperatures. Taking these into account, we think the fitting strategy using the low- and high-temperature data with the results shown as the solid line is more reasonable. As for the suppression of the thermal conductivity from 40 to 200~K as compared to the fitting, we believe that it actually indicates extra scattering mechanisms such as magnon-phonon scattering need to be considered.}

\begin{figure}[htb]
\centering
\includegraphics[width=1.0\linewidth]{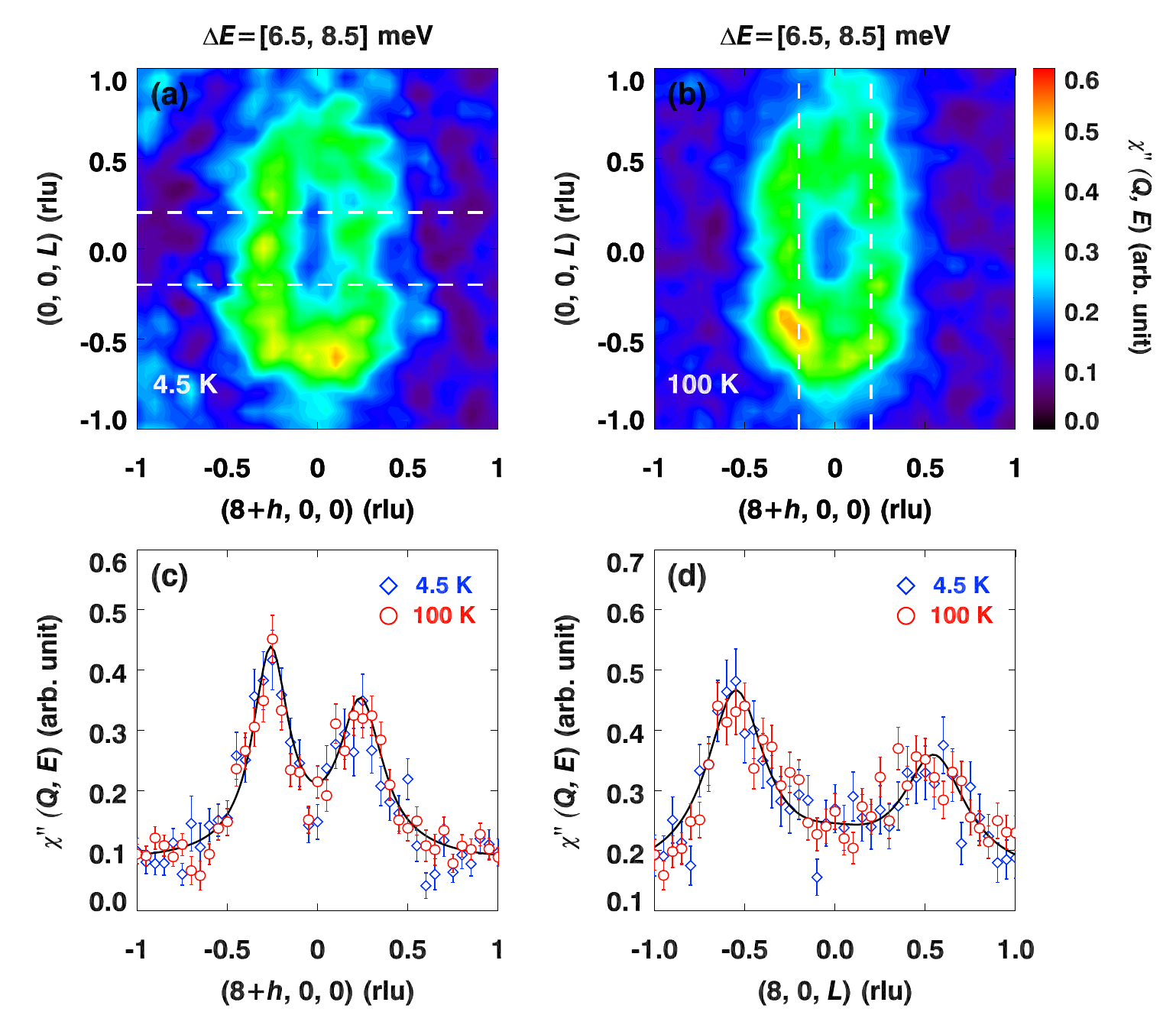}
\caption{\label{fig:ecuts}{(a) and (b) Constant-energy contours around (8,\,0,\,0) measure at 4.5 and 100~K, respectively, both with energies integrated over [6.5, 8.5]~meV and another wave vector $K$ integrated over [-0.2, 0.2]~rlu. (c) and (d) Scans through (8,\,0,\,0) along the [100] and [001] directions, respectively, with an interval of $\pm0.2$~rlu about the center, as indicated by the dashed rectangles in (a) and (b). Solid lines in (c) and (d) are the fits to the data with Lorentzian functions. Errors represent one standard deviation throughout the paper.}}
\end{figure}

\subsection{INS spectra}

The thermodynamic results indicate that while the thermal transport is dominated by phonons, magnons and phonons couple in some way and affect the overall thermal properties. To figure out the lattice dynamics and the underlying origin of the magnon-phonon scattering in \cto, we performed INS measurements. Neutron scattering can directly measure the scattering function $S({\bm Q},E)$, which is proportional to the imaginary part of the dynamical susceptibility $\chi''({\bm Q},E)$ through\cite{Shirane2002}
\begin{equation}\label{scattering}
   S({\bm Q},E)\propto\frac{\chi''({\bm Q},E)}{1-{\rm e}^{-E/k_{\rm B}T}},
\end{equation}
where $E$ and ${\bm Q}$ represent the energy and momentum transfers, respectively, and $(1-{\rm e}^{-E/k_{\rm B}T})^{-1}$ is the Bose factor. The scattering function contains both magnon and phonon information. In \cto, since magnetic order does not enlarge the size of its primitive unit cell, it is difficult to distinguish the magnon and the phonon excitations according to their different locations in the momentum space. However, taking advantage of their different dependences of intensities against ${\bm Q}$, {\it i.e.}, intensities increase as a function of $|{\bm Q}|^{2}$ for phonons but decrease for magnons, we can probe the phonon excitations at large ${\bm Q}$s. Also to eliminate the influence of Bose statistics, the measured neutron scattering intensities are divided by the Bose factor throughout the paper.

\begin{figure}[htb]
\centering
\includegraphics[width=1.0\linewidth]{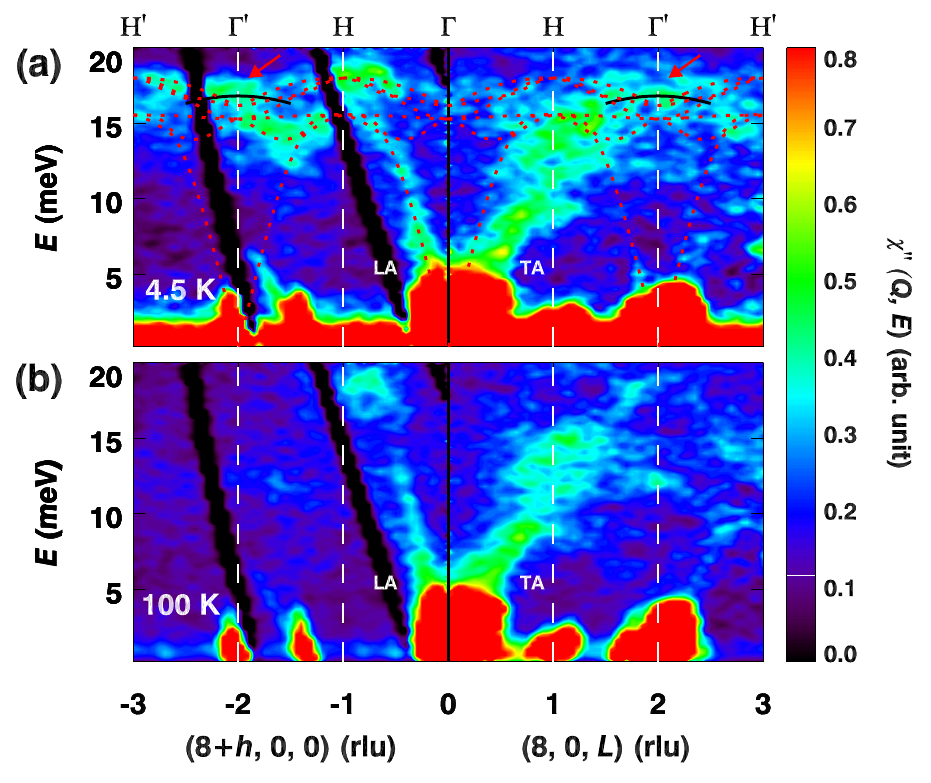}
\caption{\label{fig:dispersions}{(a) and (b) INS results for the excitation spectra of Cu$_{3}$TeO$_{6}$ measured at 4.5, 100~K, respectively. The left and right panels are the excitations propagating along the [100] and [001] directions, respectively. The integration intervals of the other two wave vectors are both chosen as $\pm0.2$~rlu. The dashed lines are the results of the linear spin-wave calculations, based on the isotropic $J_1$-$J_9$ model described in Ref.~\onlinecite{Yao2018}. Solid lines are guides to the eye to illustrate the slight downward dispersion of the additional mode, {\new and arrows indicate the positions of the mode}. Vertical dashed lines indicate high-symmetry points in the Brillouin zone.}}
\end{figure}

In Fig.~\ref{fig:ecuts}(a), we plot the constant-energy contour around the large ${\bm Q}$ of (8,\,0,\,0) at 4.5~K, with the energy interval of $7.5\pm1$ meV. They are phonon excitations and differ from magnetic excitations in two aspects: First, the excitations exist at 100~K, $\sim$40~K above $T_{\rm N}$ as shown in Fig.~\ref{fig:ecuts}(b), while magnetic excitations get featureless above $T_{\rm N}$ (Ref.~\onlinecite{Bao2018}).  Second, the contours at both temperatures show elliptical shapes centered at (8,\,0,\,0), with their long and short axes along symmetrically equivalent [001] and [100] directions, respectively, indicating anisotropic propagations of the excitations. These two features can also be seen in the $\bm{Q}$ scans shown in Fig.~\ref{fig:ecuts}(c) and (d). In each panel, the data of 4.5 and 100~K are almost identical after the Bose factor correction, confirming the phononic behaviors. The peaks along the [100] direction [Fig.~\ref{fig:ecuts}(c)] are much closer to the center than those along the [001] direction [Fig.~\ref{fig:ecuts}(d)]. Such anisotropic propagations are related to the phonon polarization. In fact, near the Bragg peak (8,\,0,\,0), neutron scattering only probes those phonon excitations with the [100] polarization. Therefore, the phonons propagating parallel and perpendicular to the [100] direction are defined as longitudinal and transverse phonons, respectively.  For acoustic phonons, the longitudinal (LA) modes have larger velocities than those of transverse (TA) modes.

\begin{figure}[htb]
\centering
\includegraphics[width=1.0\linewidth]{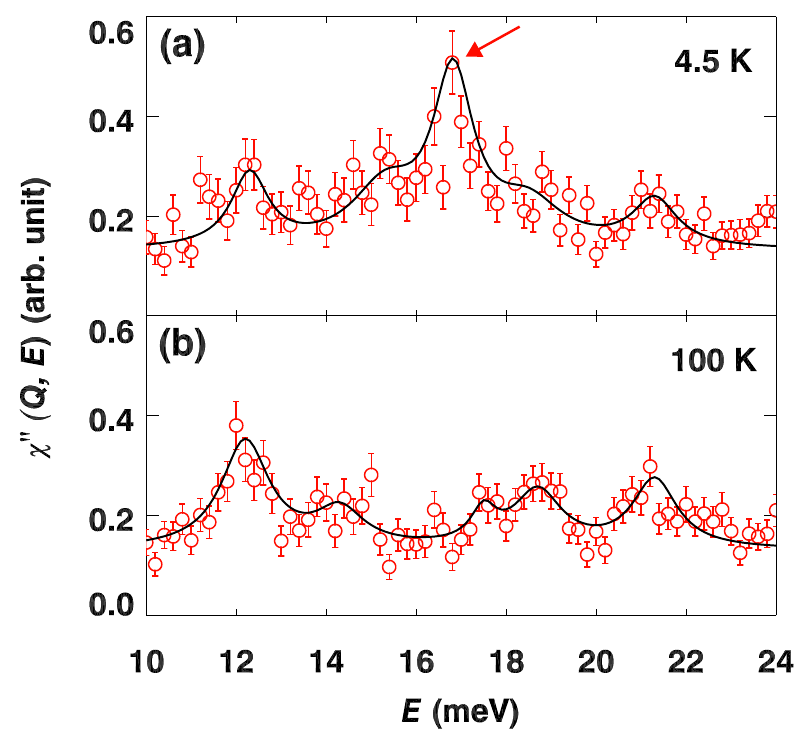}
\caption{\label{fig:qcuts}{(a) and (b) Constant-${\bm Q}$ cuts at (8,\,0,\,2) at 4.5 and 100~K, respectively. The integration intervals for $H$, $K$ and $L$ are all $\pm0.2$~rlu. Solid lines are the fits with Lorentzian functions, {\new and the arrow indicates the position of the magnon-polaron mode}.}}
\end{figure}

After orientating the phonon excitations in the momentum space, we map out the phonon excitation spectra dispersing up from (8,\,0,\,0) at 4.5 and 100~K in Fig.~\ref{fig:dispersions}. The left and right panels show the phonons propagating along the [100] and [001] directions, respectively. Although these spectra are mapped for several Brillouin zones, there are only intensive acoustic modes around (8,\,0,\,0). The observed LA and TA phonons labeled in Fig.~\ref{fig:dispersions} have different slopes, from which the velocities can be obtained as 6900 and 3150~m/s for the LA and TA phonons, respectively. The averaged velocity $\bar{v}$ can be approximated as $3/{\overline {v}^3}=1/{v_{\rm LA}^3}+2/{v_{\rm TA}^3}$ (Ref.~\onlinecite{Li2002} ), where $v_{\rm LA}$ and $v_{\rm TA}$ are the velocities for the LA and TA modes, respectively. With this, we obtain a value of 3600~m/s for $\bar{v}$. This value is comparable to the 3900~m/s obtained from the thermodynamic results. The acoustic phonons do not change significantly against temperatures. In contrast, the optical modes with energies around 15~meV near (6,\,0,\,0) and (8,\,0,\,2) vanish when the system is warmed up to 100~K, and thus cannot be phonon modes as are the acoustic branches. Considering the temperature dependence, they may be magnons. Therefore, on top of the experimental data, we plot the magnon dispersions along the $\Gamma$-H path as the dashed lines in Fig.~\ref{fig:dispersions}(a), based on linear spin-wave calculations with isotropic $J_1$-$J_9$ model described in Ref.~\onlinecite{Yao2018}. Overall, the spin-wave calculations agree well with the experimental spectra for the optical branches, as shown in Fig.~\ref{fig:dispersions}(a) and Refs~\onlinecite{Yao2018,Bao2018}, indicating the magnon origin. However, we find that the optical modes around (6,\,0,\,0) deviate from the magnon excitations. Especially, there occurs an additional mode around 16.8~meV, illustrated by the solid line in Fig.~\ref{fig:dispersions}(a). It exhibits a slight downward dispersion around (6,\,0,\,0), different from the upward shape of the calculated spin waves. This mode can also be recognized around a larger ${\bm Q}$ at (8,\,0,\,2) in the right panel of Fig.~\ref{fig:dispersions}(a). {\new By comparing the 4.5-K data in Fig.~\ref{fig:dispersions}(a) and 100-K data in Fig.~\ref{fig:dispersions}(b), it is clear that this mode is absent at 100~K.}

{\new To better reveal the temperature dependence of the additional mode observed, the constant-${\bm Q}$ cuts at (8,\,0,\,2), where the intensities of optical magnons further decrease, are plotted in Fig.~\ref{fig:qcuts}. At 100~K, there are five optical phonon modes that can be resolved in the selected energy window. Among them, the modes at 12.2, 14.3 and 21.3~meV are close to the 93.5, 120 and 169~${\rm cm}^{-1}$ Raman-active zone center optical phonons observed in Raman measurements\cite{Choi2008}. At 4.5~K, except the modes at 12.2 and 21.3~meV, other phonon modes are not well resolved, due to the emergence of the optical modes of the magnetic excitations. On top of this, there is a mode manifested as a strong peak centered at 16.8~meV [Fig.~\ref{fig:qcuts}(a)]. This mode is absent at 100~K as shown in Fig.~\ref{fig:qcuts}(b), where the intensities around 16.8~meV are on the background level.}  

\section{Discussions}

What is the origin of the mode at 16.8~meV around the zone center shown in Figs~\ref{fig:dispersions}(a) and \ref{fig:qcuts}(a)? Since this mode disappears at high temperatures, it cannot be a phonon. We do not think it is a magnon either because: i) {\new This mode is absent around the $\Gamma$ point at low $\bm{Q}$s as shown in Ref.~\onlinecite{Bao2018} where topological magnons were observed at low $\bm{Q}$s.} If it were a magnon, the intensity would be much stronger as of other magnons due to the magnetic form factor. {\new In fact, the high-$\bm Q$ data obtained from the experiment that led to the work in Ref.~\onlinecite{Bao2018} did reveal some hints on this mode, which motivated us to carry out more detailed INS studies with doubled sample mass in the current work.} ii) This mode is not reproduced by the linear spin-wave calculations using the $J_1$-$J_9$ model. As demonstrated in Ref.~\onlinecite{Yao2018}, this model can fit the experimental magnon spectra quite well, but this mode with a downward dispersion appears to deviate from the calculated upward spin-wave excitations. iii) As shown in Fig.~\ref{fig:qcuts}(a), the intensity of this mode is much stronger than those of the magnons, which are too weak to be resolved from the phonons. Instead, we propose it to be a magnon-polaron mode, which is hybridized excitations between magnon and phonons due to the magnon-phonon coupling.

The magnon-polaron mode can be generated around the anticrossing region where a gap will be opened at the intersection of the magnon and phonon dispersions due to magnon-phonon coupling\cite{PhysRevB.91.104409,PhysRevB.92.214437}. For example, in YIG (Refs~\onlinecite{PhysRevLett.117.207203,PhysRevB.96.100406}) and YMnO$_3$ (Refs~\onlinecite{PhysRevLett.99.266604,PhysRevB.97.134304}), hybridizations take place at the anticrossings between the acoustic magnon and acoustic phonon branches. While for some materials with acoustic magnons having much larger velocities than acoustic phonons, such as CuBr$_{2}$ (Ref.~\onlinecite{PhysRevB.96.085111}) and Mn$_{3}$Ge (Ref.~\onlinecite{PhysRevB.99.214445}), the anticrossings occur between acoustic magnon and optical phonon branches. However, in this material \cto, the approximately linear dispersions approaching to the zone center with similar velocities for both acoustic magnons and LA phonons [Fig.~\ref{fig:dispersions}(a)] forbid the formations of the  hybridized magnon-polaron mode in the two ways mentioned above. We think in our case, this mode is due to the  hybridizations of the optical magnon and optical phonon branches, that are very close to the zone center. This can explain the 16.8-meV mode near the zone center we observed in Figs~\ref{fig:dispersions} and \ref{fig:qcuts}. In fact, earlier Raman measurements observed an anomalous mode at  132~${\rm cm}^{-1}$~(Ref.~\onlinecite{Choi2008}). This value is almost the same as the 16.8~meV we observe here.

The occurrence of the hybridized mode between magnons and phonons indicates the magnon-phonon coupling in this system, which is what our thermodynamic results also imply. It is worth noting that phonons are the dominant carriers of heat current in \cto [Fig.~\ref{fig:thermal}(b)], which is different from other topological magnonic materials such as Lu$_{2}$V$_{2}$O$_{7}$ (Refs~\onlinecite{Onose297,PhysRevB.85.134411}) and Cu(1,3-bdc) (Refs~\onlinecite{PhysRevLett.115.147201,PhysRevLett.115.106603}), where magnons can also act as carriers of heat current and even exhibit thermal Hall effect\cite{Onose297,PhysRevLett.115.106603}. With the conventional kinetic gas theory, the thermal conductivity can be approximated by $\frac{1}{3}C_{\rm ph}v_{\rm ph}l_{\rm ph}+\frac{1}{3}C_{\rm m}v_{\rm m}l_{\rm m}$ (Ref.~\onlinecite{PhysRevB.90.064421}). The $v_{\rm ph(m)}$ and $l_{\rm ph(m)}$ are velocities and mean free paths for phonons (magnons), respectively. In \cto, the magnetic specific heat is analyzed to have negligible contributions to the total specific heat (Eq.~\ref{HC}). Provided similar $v_{\rm m}$ and $l_{\rm m}$ with $v_{\rm ph}$ and $l_{\rm ph}$ respectively, the contribution of magnons to the thermal conductivity is also negligible. Nevertheless, magnons can still manifest themselves by scattering phonons\cite{PhysRevLett.93.177202,PhysRevB.8.2130}, that may explain the deviation from 40 to 200~K between the measured data and the fitting with Debye-Callaway model in Fig.~\ref{fig:thermal}(b). Above $T_{\rm N}$, phonons will be scattered by the spin fluctuations although there are no well-defined magnons. With decreasing temperature, the fluctuations will be suppressed and the system orders antiferromagnetically below $T_{\rm N}$. In the ordered state, phonons are scattered by magnons due to the magnon-phonon coupling. The combination of these two scattering mechanisms will cause the suppression of the phonon thermal conductivity in Fig.~\ref{fig:thermal}(b). Similar situation was also found in YMnO$_3$~(Ref.~\onlinecite{PhysRevLett.93.177202}), where suppression of thermal conductivity up to at least 300~K was reported\cite{PhysRevLett.93.177202}. However, compared to YMnO$_3$, the suppression we observe here is smaller. We believe this is because the magnon-phonon coupling here mostly affects the optical phonons and magnons, as opposed to  YMnO$_3$ where the hybridized magnon-polaron mode occurs around the region of anticrossing between the acoustic magnon and phonon branches\cite{PhysRevLett.99.266604,PhysRevB.79.134409}, and has larger contributions to the thermal conductivity. For the magnon-polaron mode at such a high energy of 16.8~meV in our case, we think its suppression of the thermal conductivity is quite small, especially at low temperatures. Nevertheless, with the magnon-phonon coupling suggested by the suppression, a magnon-polaron mode can emerge when magnons and phonons are to intersect.  

\section{Summary}

In summary, our specific heat and thermal conductivity measurements indicate phonons are the dominant carriers of heat current in \cto. Magnons manifest themselves by scattering phonons, causing the suppression of thermal conductivity around $T_{\rm N}$. Our INS measurements find an additional mode located at about 16.8~meV, and we attribute it to a magnon-polaron mode. We consider the suppression of thermal conductivity and the emergence of the magnon-polaron mode to be evidence for magnon-phonon coupling in \cto. Considering that the new mode is close to the linear magnon band crossings\cite{Bao2018,Yao2018}, we think this material provides a new platform for studying the interplay between topological magnons and other emergent excitations.

\section{Acknowledgements}

We would like to thank Jianfei Gu and Xi Zhang at Nanjing University for stimulating discussions, and Long Tian and Prof. Xingye Lu at Beijing Normal University for allowing us to use their Laue machine. The INS experiment at MLF of J-PARC was performed under a user program (Proposal No.~2018B0036). The work was supported by National Natural Science Foundation of China with Grant Nos~11822405, 11674157, 11674158, 11774152, 11904170, 11974162 and 11834006, Natural Science Foundation of Jiangsu province with Grant Nos~BK20180006 and BK20190436, Fundamental Research Funds for the Central Universities with Grant No.~020414380117, Innovative and Creative Research Program for Doctoral Students of Nanjing University with Project No.~CXCY18-13, and the Office of International Cooperation and Exchanges of Nanjing University.


%

\end{document}